\documentclass[a4paper]{article}

\usepackage{INTERSPEECH2022}
\usepackage[dvipsnames]{xcolor}
\usepackage{multicol}
\usepackage{multirow}
\usepackage{hyperref}

\title{Decision Attentive Regularization to Improve Simultaneous Speech Translation Systems}
\name{Mohd Abbas Zaidi$^{\star}$\thanks{$^{\star}$Equal contribution. Accepted at Interspeech 2022}, Beomseok Lee$^{\star}$, Sangha Kim, Chanwoo Kim}
\address{
  Samsung Research, Seoul, South Korea}
\email{\{abbas.zaidi, bsgunn.lee, sangha01.kim, chanw.com\}@samsung.com}

\begin{document}

\maketitle
\begin{abstract}
  Simultaneous translation systems start producing the output while processing the partial source sentence in the incoming input stream. These systems need to decide when to \textit{read} more input and when to \textit{write} the output. These decisions depend on the structure of source/target language and the information contained in the partial input sequence. Hence, \textit{read/write} decision policy remains the same across different input modalities, i.e., speech and text. This motivates us to leverage the text transcripts corresponding to the speech input for improving simultaneous speech-to-text translation (SimulST). We propose Decision Attentive Regularization (DAR) to improve the decision policy of SimulST systems by using the simultaneous text-to-text translation (SimulMT) task. We also extend several techniques from the offline speech translation domain to explore the role of SimulMT task in improving SimulST performance. Overall, we achieve 34.66\% / 4.5 BLEU improvement over the baseline model across different latency regimes for the MuST-C English-German (EnDe) SimulST task.
\end{abstract}

\section{Introduction}
\label{sec:intro}

\begin{figure}[t]
     \centering
    %  \vspace{-0.85cm}
    \hspace{-0.85cm}

     \includegraphics[width=6.5cm, height=12cm]{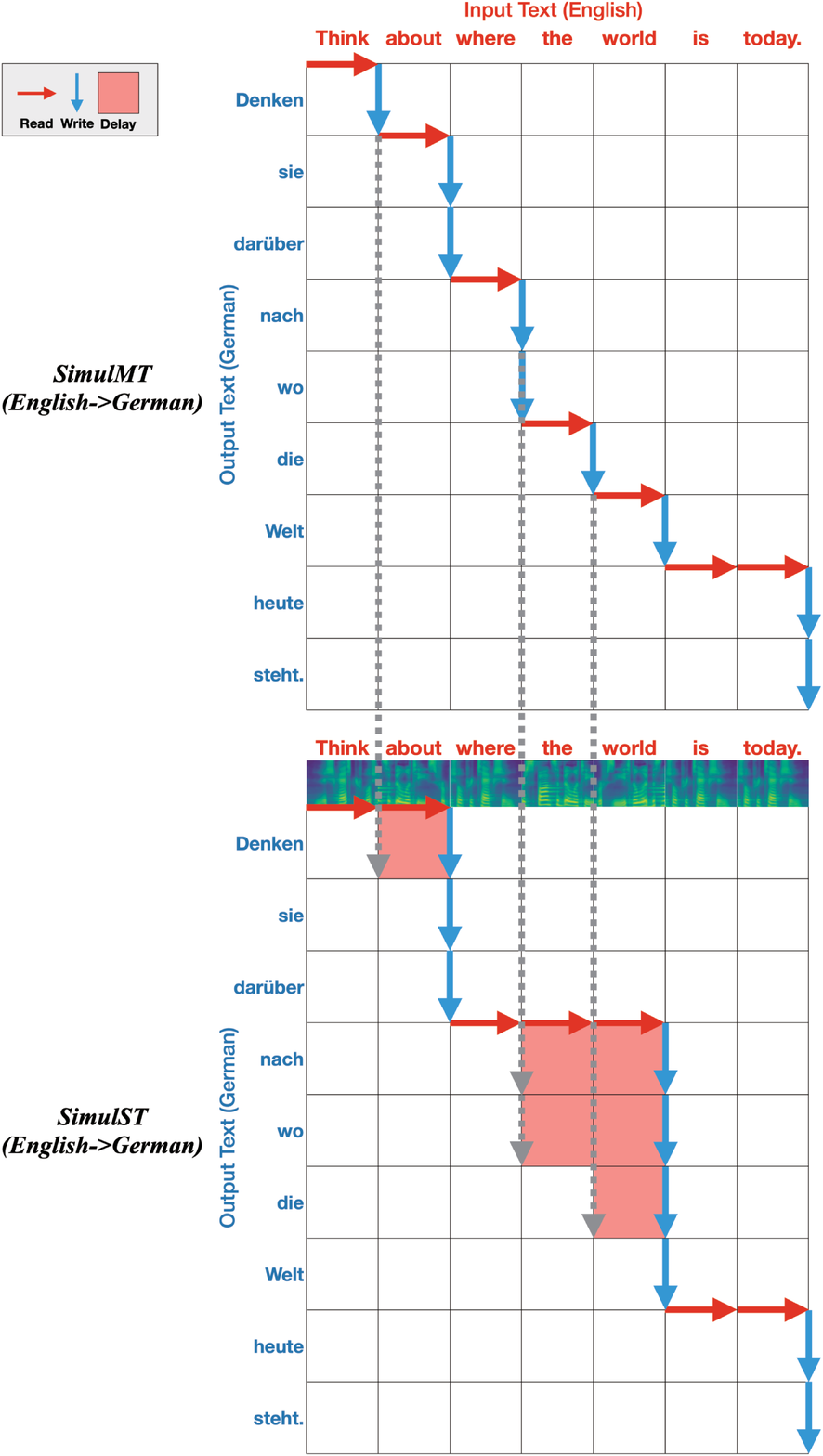}
     \caption{An example of using SimulMT decision to improve SimulST for English-German translation. SimulMT can prevent the SimulST model from incurring extra delays.}
     \label{fig:approach}
    %  \vspace{-0.3cm}
 \end{figure}

\subsection{Simultaneous Translation} Simultaneous translation systems find huge applications in real life scenarios such as live subtitle generation and real-time interpretation. To provide translation in tandem with the streaming input, these systems alternate between \textit{read/write} decisions, i.e., \textit{reading} the source sequence and \textit{writing} the target tokens. Initial approaches for such systems, such as wait-$k$ \cite{ma2018stacl} use a fixed policy, where the \textit{read/write} schedule is pre-decided. Recent works \cite{arivazhagan2019monotonic, ma2019monotonic} use monotonic attention \cite{raffel2017online, chiu2017monotonic} to learn a flexible decision policy. Monotonic attention provides a closed form expression for the expected input output alignment to train the discrete \textit{read/write} decisions in expectation. Monotonic multihead attention (MMA) \cite{ma2019monotonic} replaces the soft attention in the transformer \cite{Transformer} model with monotonic attention outperforming the models with fixed decision policy. 
% \vspace{-0.2cm}

\subsection{Offline Speech Translation} The offline speech translation (ST) task has traditionally suffered from data scarcity issues. Hence, various approaches use pretraining \cite{bansal2018pre}, multitask learning \cite{our_icassp20, tang2021improving}, meta-learning \cite{han2020end} and knowledge distillation (KD) \cite{liu2019end} to leverage the high resource machine translation (MT) and automatic speech recognition (ASR) tasks for improving ST performance. A recent work \cite{tang2021improving} improves information sharing between offline MT and ST task by using online KD and cross-attentive regularization (CAR). These approaches have demonstrated that the MT task can be crucial to improve the performance of ST systems.
% \vspace{-0.2cm}

\subsection{Simultaneous Speech Translation} Simultaneous speech translation (SimulST) systems perform realtime speech-to-text translation. A recent work \cite{ma2020simulmt} brings together the advances in transformer \cite{Transformer} based offline ST systems and montonic multihead attention (MMA) based simultaneous translation. 

For MMA-based SimulST model, \textit{read/write} decisions are guided by the monotonic attention energies learned during training. In the absence of direct supervision for the decision policy, the MMA model learns these decisions by balancing the trade-off between output accuracy and latency. These \textit{read/write} decisions depend on the amount of information contained in the source sequence, and the word orders of the source/target languages. Hence, the decision policy for speech and text inputs remains the same. Moreover, due to the relative complexity associated with speech inputs, it is easier to learn the \textit{read/write} decisions from the text data. Hence, SimulMT decisions can potentially serve as a reference to improve the SimulST decision policy as shown in Figure \ref{fig:approach}. 

We propose Decision Attentive Regularization (DAR) which utilizes the monotonic attention energies of the SimulMT model to guide the decision policy of the SimulST model implicitly. We also extend several techniques from the offline ST domain, such as multitask learning, online KD and CAR to the SimulST task. Experiments on the MuST-C EnDe dataset show that the proposed DAR, along with other approaches improves the performance of MMA-based SimulST systems significantly. 

\section{MODEL}
\label{sec:model}

We use the MMA model described in \cite{ma2020simulmt} as our baseline SimulST model. It processes partial input speech and partial target text to produce the next target token. Given partial source and target ($x_{\leq j}  \in \mathbf{x}, y_{< i} \in \mathbf{y}$), the next target token $y_{i}$ is generated as follows:
\begin{gather}
    h_{j} = \mathcal{E}(x_{\leq j}) \label{eq:xr} \\
    s_i = \mathcal{D}(y_{<i}, \mathcal{MA}(s_{<i},h_{\leq j})) \label{eq:yr} \\
    y_i = Output(s_i) \label{eq:output}
\end{gather}

\noindent where $\mathcal{E}(.)$ and $\mathcal{D}(.)$ represent the encoder and decoder blocks and $\mathcal{MA}$ refers to the monotonic multihead attention energy function. As seen in Figure \ref{fig:approach}, the MMA model alternates between \textit{read} and \textit{write} decisions at test time. It uses the monotonic attention energies to make \textit{read/write} decisions as follows:
\begin{gather}
    e_{i,j} = \mathcal{MA}(s_{i-1}, h_{j}) \label{eq:me} \\
    p_{i,j} = Sigmoid(e_{i,j}) \label{eq:rwd} \\
    z_{i,j} \sim Bernoulli(p_{i,j}) \label{eq:pd}
\end{gather}

When $z_{i,j}=1$ (\textit{write} decision), the model sets $t_i=j$ and computes the decoder output using $h_{\leq j}$, where $t_i$ refers to the number of encoder states required to produce the $i_{th}$ decoder output. If $z_{i,j}=0$, the model needs to read further. It computes $h_{j+1}$ and repeats Eq. \ref{eq:me} to \ref{eq:pd}.

As mentioned earlier, the proposed DAR loss aims to improve the SimulST decision policy using the monotonic energy activation of the SimulMT model.
% as depicted in Figure \ref{fig:architecture}. 
DAR computes the similarity between the monotonic energies of speech and text input corresponding to each training example. However, it cannot be computed directly since the attention energies corresponding to speech ($\mathbf{A}^s$) and text ($\mathbf{A}^t$) have different sizes due to different input lengths. Similar to \cite{tang2021improving}, we use self-attention and cross-attention operations with respect to $\mathbf{A}^t$ to obtain attention representations $\mathbf{A}^{t\rightarrow t}$ and $\mathbf{A}^{s\rightarrow t}$ which have the same size. Finding the $\mathcal{L}2$ distance between these reconstructed representations provides the required cross-modal similarity metric for each example. 

During joint training of SimulST and SimulMT, each training example consists of input speech and corresponding transcript in the source language and output text in the target language. Let K and L denote the length of speech and text representations at the output of the encoder. The monotonic attention energy matrices for speech and text for the $h_{th}$ head are defined as follows:
% \vspace{-0.1cm}
% \begin{gather}
%     \mathbf{A}^s_h=(att_{h,1}^{s}, att_{h,2}^{s}, \ldots ,att_{h,K}^{s}) 
%     \label{eq:ab} \\

%     \mathbf{A}^t_h=(att_{h,1}^{t}, att_{h,2}^{t}, \cdot \cdot \cdot,att_{h,L}^{t})   
%     \label{eq:ac}
% \end{gather}
\vspace{-0.2cm}
\begin{align*}
    \mathbf{A}^s_h = (at_{h,1}^{s}, \cdot \cdot ,at_{h,K}^{s}), \mathbf{A}^t_h = (at_{h,1}^{t},\cdot \cdot,at_{h,L}^{t})
    % p_{i,j} = Sigmoid(e_{i,j}) \label{eq:rwd} \\
    % z_{i,j} \sim Bernoulli(p_{i,j}) \label{eq:pd}
\end{align*}
% \vspace{-0.2cm}
\noindent where $at_{j} = e_{(i,:)}$ refers to the monotonic attention corresponding to the $j_{th}$ encoder output token, aggregated across all decoder indices. Attention energies from $H$ different heads are stacked as follows:
% \vspace{-0.4cm}
\begin{equation}
% \label{eq:enc-output}
% \setstretch{0.6}
\mathbf{A}^s=[ \mathbf{A}_{1}^{s}, \ldots,\mathbf{A}_{H}^{s}] = (a_{1}^{s}, a_{2}^{s}, \ldots,a_{K\times H}^{s})
% \mathbf{A}^t=[ \mathbf{A}_{1}^{t}, \mathbf{A}_{2}^{t}, \cdot \cdot \cdot,\mathbf{A}_{L}^{t}] \label{2}
\end{equation}
% \vspace{-0.6cm}
\begin{equation}
% \label{eq:enc-output}
% \mathbf{A}^s=[ \mathbf{A}_{1}^{s}, \mathbf{A}_{2}^{s}, \cdot \cdot \cdot,\mathbf{A}_{L}^{s}]
\mathbf{A}^t=[ \mathbf{A}_{1}^{t},  \ldots,\mathbf{A}_{H}^{t}] = (a_{1}^{t}, a_{2}^{t}, \ldots ,a_{L\times H}^{t})
\end{equation}

\noindent Next, similarity matrix $\mathbf{S}$ is used to obtain $\mathbf{A}^{s\rightarrow t}$ via cross-attention between $\mathbf{A}^s$ and $\mathbf{A}^t$. Similarly, $\mathbf{A}^{t\rightarrow t}$ is obtained from $\mathbf{A}^t$ by using self-attention.

\begin{equation}
\label{eq:cos-dist}
s_{i,j}=\frac{a_{i}^{s}\cdot a_{j}^{t}}{\left \|a_{i}^{s}\right \|_2 \left \|a_{j}^{t}\right \|_2}
\end{equation}
% \vspace{-0.1cm}
\begin{equation}
\label{eq:reconstruction}
\mathbf{A}^{s\rightarrow t}=\mathbf{A}^s\cdot \text{softmax}(\mathbf{S}) 
\end{equation}
\noindent For the $d_{th}$ decoder layer, the DAR loss is computed as follows: 
\begin{equation}
\label{eq:dar-loss}
\mathcal{L}_{DAR}^d(\theta_s)= \sum \frac{1}{LH}\left \| \mathbf{A}^{s\rightarrow t} - sg[\mathbf{A}^{t\rightarrow t}] \right \|_2 \\
\end{equation}
where $sg$ (stop gradient operator) allows the model to use text attention as a reference for the speech attention. Finally, the DAR loss is computed by averaging across M decoder layers.
\begin{equation}
\mathcal{L}_{DAR}(\theta_s) = \sum_{d=1}^M \frac{1}{M} \mathcal{L}_{DAR}^d(\theta_s) 
\end{equation}

In addition to the proposed DAR approach, this work also extends several existing techniques from the offline ST domain. Firstly, it employs multitask learning by training SimulMT model along with SimulST. It also extends online KD and CAR \cite{tang2021improving} losses to SimulST. Finally, similar to other MMA-based translation systems, it uses differentiable average lagging (DAL) \cite{dal} loss to control the latency of simultaneous translation models. The overall loss $\mathcal{L}(\theta_s, \theta_t)$ is defined as follows:
\begin{equation}
\begin{aligned}
\label{eq:loss}
\mathcal{L}(\theta_s, \theta_t) = & \; (1-\alpha)\mathcal{L}_{ST-NLL}(\theta_s) +  \alpha\mathcal{L}_{KD}(\theta_s, \theta_t)  \\
& + \beta\mathcal{L}_{CAR}(\theta_s) + \gamma\mathcal{L}_{MT-NLL}(\theta_t) \\ 
& + \delta \mathcal{L}_{DAR}(\theta_s, \theta_t)+  \lambda\mathcal{L}_{DAL}(\theta_s, \theta_t)\\
\end{aligned}
\end{equation}

\noindent It combines the negative-log likelihood loss for both speech (ST) and text (MT) with KD, CAR, DAR, and DAL loss. 
($\theta_s$, $\theta_t$: speech/text model parameters)

\section{EXPERIMENTAL SETTINGS}
\label{sec:exp}

\subsection{Dataset} For SimulST, MuST-C \cite{di2019must} English-German (En-De) dataset is used for training with tst-COMMON as the test set. For SimulMT, WMT 14 \cite{bojar2014findings} and MuST-C En-De serve as the training data. Table \ref{table:data} provides the dataset statistics. Data preprocessing details are the same as \cite{tang2021improving}. 
% \vspace{0.1cm}
\begin{table}[hbt!]
\hspace{0.6cm}
% \centering
\begin{tabular}{ |c|c|c|c|c|}
 \hline
 \multirow{2}{*}{Task} & \multirow{2}{*}{\# Hours}  & \multicolumn{3}{c|}{\# Sentences}   \\  \cline{3-5} 
 
 \multirow{2}{*}{} & \multirow{2}{*}{}  & Train & Dev & Test     \\ 
 \hline \hline
 MuST-C & 408 & 225k & 1,423 & 2,641   \\
 WMT 14 & - & 2.57M & 26k & 3003  \\
 \hline

\end{tabular}

\caption{Dataset Statistics (\small{\# - Number of})}
\label{table:data}
\vspace{-0.5cm}
\end{table}

\subsection{Data Augmentation} Equal amounts of synthetic speech and text data is generated for MuST-C EnDe dataset. Augmented speech is paired with original text and vice versa. Synthetic speech is generated by varying the SoX effects similar to \cite{han2020end}, while the augmented target text is generated by translating \cite{backtranslation} the source transcripts using the WMT 19 winner offline MT En-De model \cite{ng2019facebook}.

\subsection{Pretraining and Weight Sharing}
\label{pret}
% \noindent \textbf{Pretrained models}
We use MMA-based transformer as our base model. The transformer decoder of the offline ST model \cite{tang2021improving} is replaced with a monotonic decoder \cite{ma2020simulmt}. The base model consists of a speech encoder (12 layers), text encoder (6 layers), and a joint monotonic decoder (6 layers) shared between the SimulST and SimulMT models. Similar to \cite{tang2021improving}, the top 6 layers of the speech encoder are tied to the text encoder. The speech encoder is initialized using pretrained ASR encoder\footnote{ASR Model: \url{https://dl.fbaipublicfiles.com/fairseq/s2t/mustc_joint_asr_transformer_m.pt}}. MT encoder and joint decoder are initialized using an offline MT \footnote{MT Model: \url{https://dl.fbaipublicfiles.com/joint_speech_text_4_s2t/must_c/en_de/checkpoint_mt.pt}} model.

\begin{table}[t]
\hspace{0.5cm}
% \centering
\begin{tabular}{ c c c c}
 \hline
\multirow{2}{*}{Hyper-parameter} & \multirow{2}{*}{Value} \\
%  Hyperparameter &  &  
% maybe remove same value?
\\ \hline \hline
 speech conv layers & 2 \\
 speech conv stride & (2,2) \\
%  speech encoder layers & 12 \\
%  text encoder layers & 6 \\
 shared encoder layers & 6 \\
 encoder embed dim & 512 \\
 encoder ffn embed dim & 2048 \\
 encoder attention heads & 8 \\
%  decoder layers & 6 \\
 decoder embed dim & 512 \\
 decoder ffn embed dim & 2048 \\
 decoder attention heads & 8 \\
 dropout & 0.1 \\
 optimizer & adam  \\
 adam-$\beta$ & (0.9, 0.999) \\
 clip-norm & 10.0 \\
 lr scheduler & inverse sqrt \\
 learning rate & 0.002 \\
 warmup-updates & 20000 \\
 label-smoothing & 0.1 \\
 max text tokens per batch & 5000 \\
 max speech frames per batch & 5000 \\
%  latency weight average & (0.01,0.05,0.1) \\
%  \#params     & $\approx39M$ & $\approx39M$ \\
 
 \hline
\end{tabular}
\caption{List of Hyperparameters}
\label{table:common_hparams}
\vspace{-0.5cm}
\end{table}

\subsection{Training Details} All the models are implemented using the Fairseq \cite{ott2019fairseq} toolkit. In order to compute the DAR loss, parallel ST and MT data is required. The speech transcripts in the MuST-C EnDe dataset are utilized as the parallel SimulMT input required to compute the cross-modal similarity losses.

All the models are trained using 8xA100 GPUs with an update frequency of 4. The training schedule is the same as \cite{ma2020simulmt}. The model is first trained for 150 epochs (110 hours / 4.5 days) without the differentiable average lagging (DAL) latency loss by setting $\lambda=0$ in Eq. \ref{eq:loss}. These models are referred to as $\lambda_0$ models. Finally, the models are finetuned further for 50 epochs (40 hours) after adding the DAL loss. The training with the latency loss is carried out using three different values of $\lambda \in \{0.01,0.05,0.1\}$. During inference, the step sizes are varied ($\{120,200,280,360,440,520\}$ in ms) to obtain the model performance in different latency regimes. Speech-segment size / step size refers to the duration of speech consumed corresponding to each \textit{read} decision. The weights for various losses in Eq. \ref{eq:loss} are as follows: $\alpha = 0.2$ for online KD, $\beta =0.02$ for CAR and $\gamma=0.5$ for MT-NLL. $\delta=0.01$ is the weight for the proposed DAR loss. It was obtained using standard grid-search from 0.05 to 0.5 with a step-size of 0.05. Detailed hyperparameter settings required to replicate the results can be found in the Table \ref{table:common_hparams}.

\section{RESULTS}
\label{sec:results}

As mentioned in Section \ref{sec:model} and \ref{sec:exp}, this work extends data augmentation, multitask learning, online KD and CAR based techniques from offline ST domain to the SimulST task. It also proposes a new DAR loss to SimulST systems. Table \ref{table:res} reports the performance and improvements for the $\lambda_0$ models trained using different methods. In order to plot the latency-quality curves \cite{ma2018stacl}, multiple models are trained with different $\lambda$ values. Further, during inference, step-sizes are varied to obtain model performance in different latency regimes. Case-sensitive detokenized BLEU \cite{post2018call} is used to measure the quality while average lagging (AL) \cite{ma2020simulmt} is used as the latency metric.

\begin{table}[hbt!]
\hspace*{-0.4cm}
\centering
\begin{tabular}{ |c|l|c|c|c|}
 \hline
 \multirow{2}{*}{No} & \multirow{2}{*}{Methods} & \multirow{2}{*}{Model Name} & \multicolumn{2}{c|}{$\lambda_0$ models}   \\
  &  & &  BLEU & $\Delta$ \\

\hline
\hline
% {\color{blue}1}
%  {\color{ForestGreen}5}
%  {\color{red}6}
 1   & Baseline &  \textit{MMA}  & 17.23 & -    \\ 
 2   & 1 + Data Augmentation & \textit{MMA-Aug}     & 19.81 & 2.58   \\
 \hline \hline
 3   &  1 + Multitask Learning & -          & 18.90 & 1.67     \\ 
 4   &    3 + Online KD & - & 19.85 & 0.95     \\
 5   &  4 + CAR \hspace{0.85cm} & \textit{MMA-MT} & 20.10 & 0.25      \\ 
 6   & 5 + Data Augmentation & \textit{MMA-Aug-MT}  & 21.49 &  1.39        \\  \hline
 \multicolumn{5}{c}{\textbf{ This Work}} \\ \hline  
 7   & 5 + \textbf{DAR} & - &  \textbf{20.67} & \textbf{0.57} \\
 8   &  6 + \textbf{DAR} \hspace{0.54cm} & \textit{MMA-DAR}    & \textbf{22.35} & \textbf{0.86} \\  \hline
\end{tabular}

\caption{Performance of various approaches: $\lambda_0$ models}
\label{table:res}
\vspace{-0.6cm}
\end{table} 
\subsection{Existing Approaches}

\subsubsection{Data Augmentation} 
\begin{figure}[t]
     \centering
    %  \vspace{-0.85cm}
     \includegraphics[width=8cm, height=6cm]{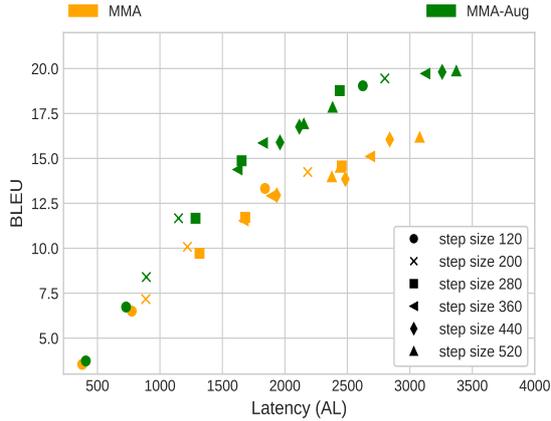}
     \caption{Effect of Data Augmentation.}
     \label{fig:aug}
    %  \vspace{-0.3cm}
 \end{figure}

As mentioned in Section \ref{sec:exp}, several data augmentation techniques are added to the MMA baseline. This model is referred to as \textit{MMA-Aug}. Data augmentation provides significant performance gains both for the baseline and our proposed approach. For $\lambda_0$ models, it improves the BLEU score by 2.58 points over \textit{MMA}. Figure \ref{fig:aug} provides the latency-quality tradeoff comparison of \textit{MMA-Aug} versus the \textit{MMA} baseline. 

\subsubsection{Multitask Learning} Multitask learning refers to simply training the SimulST and SimulMT model together with shared parameters. As seen in Table \ref{table:res}, multitask learning boosts the performance of the SimulST task by 1.67 BLEU scores for the $\lambda_0$ model.

\subsubsection{Online KD \& CAR} Similar to the offline domain, online KD and CAR improve the performance of the SimulST model as well. For the $\lambda_0$ model, online KD and CAR provide an improvement of 0.95 and 0.25 BLEU scores over the multitask learning baseline.

\begin{figure}[t]
     \centering
    %  \vspace{-0.85cm}
     \includegraphics[width=8cm, height=6cm]{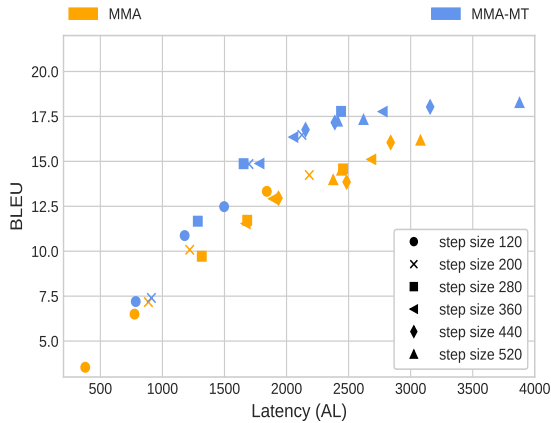}
     \caption{Effect of existing approaches related to the SimulMT task (Multitask learning, online KD \& CAR)}
     \label{fig:car}
     \vspace{-0.3cm}
 \end{figure}

All the existing techniques related to the auxillary SimulMT task such as multitask learning, online KD, and CAR are grouped together, and this model is referred to as \textit{MMA-MT}. Combined together, these approaches improve the $\lambda_0$ model performance by 2.87 BLEU score (Row 5 vs. Row 1 in Table \ref{table:res}). Figure \ref{fig:car} provides the latency-quality curves for \textit{MMA-MT} against the baseline \textit{MMA} model. It provides consistent improvements as compared to the baseline across different latency regimes. Next, another model (\textit{MMA-Aug-MT}) is trained by adding data augmentation techniques to \textit{MMA-MT}. It serves as a baseline to quantify the improvements obtain using the proposed DAR approach.

\subsection{Proposed Approach: DAR}
\label{ssec:prop}

\begin{figure}[t]
     \centering
    %  \vspace{-0.85cm}
     \includegraphics[width=8cm, height=6cm]{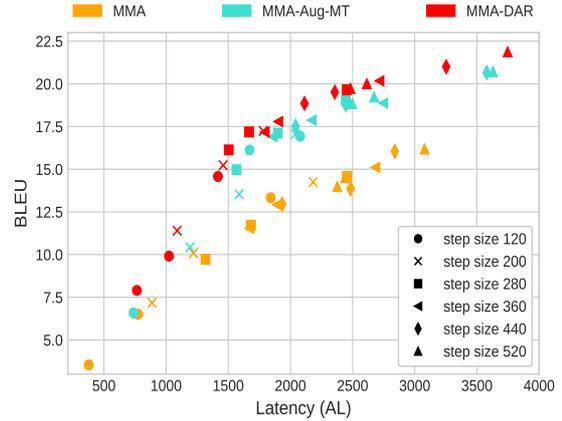}
     \caption{Effect of the proposed DAR loss.}
     \label{fig:dar}
    %  \vspace{-0.3cm}
 \end{figure}

As discussed in the previous sections, DAR loss is designed to improve the \textit{read/write} decisions for the challenging SimulST task using the relatively easier SimulMT task. For $\lambda_0$ models, DAR provides a BLEU score improvement of \textbf{0.57} over the \textit{MMA-MT} model, and \textbf{0.86} over \textit{MMA-Aug-MT}. The final model, \textit{MMA-DAR} (Row 8 in Table \ref{table:res}) is trained by using all the described approaches (Eq. \ref{eq:loss}). Figure \ref{fig:dar} provides the improvements achieved with respect to the latency-quality tradeoff using DAR. It provides improvements in the range of $0.7\sim1.2$ BLEU scores consistently across all latency regimes. It is interesting to note that the improvements obtained through decision/attention regularization are much higher as compared to regularizing the input speech encoding through CAR. 

\noindent \textbf{Validation and Overall results:} The improvements obtained from DAR are statistically significant at 99\% confidence with a p value of 0.002 (paired t-test). In addition to the latency-quality curves, we also calculate the overall performance improvement obtained using all the approaches. We choose ten different points with similar latency values and compute the absolute and relative improvements obtained by \textit{MMA-DAR} over \textit{MMA}. Table \ref{table:overall} reports the exact values used for this comparison. Each of these points is chosen such that the latency for the \textit{MMA-DAR} model is less than that of \textit{MMA}. Averaging these values, we obtain aggregated improvement of 4.5 BLEU score or 34.66\%.

\begin{table}[hbt!]
% \centering
\hspace{0.3cm}
\begin{tabular}{|c|c|c|c|c|c|}
 \hline
 \multicolumn{2}{|c|}{MMA}  &
 \multicolumn{2}{c|}{MMA-DAR}  & BLEU &  BLEU\\ \cline{1-4} 
 BLEU & AL & BLEU & AL & {$\Delta$} & {$\% \Delta$}\\
 \hline
 6.5 & 775 & 7.9 & 765 & 1.40 & 21.53 \\ 
 10.08 & 1220 & 11.40 & 1089 & 1.32 & 13.09\\
 11.72 & 1683 & 16.13 & 1504 & 4.41 & 37.63 \\
 13.33 & 1841 & 17.24 & 1781 & 3.91 & 29.33 \\
 12.92 & 1891 & 18.32 & 1794 & 5.4 & 41.79 \\
 12.95 & 1935 & 19.23 & 1902 & 6.28 & 48.94 \\
 14.24 & 2183 & 19.39 & 2113 & 5.15 & 36.17\\
 13.98 & 2376 & 19.97 & 2357 & 5.99 & 42.84\\
 13.85 & 2484 & 20.23 & 2482 & 6.38 & 46.06\\
 16.18 & 3079 & 20.97 & 2614 & 4.70 & 29.61 \\ \hline
\end{tabular}

\caption{Aggregated performance improvements}
\label{table:overall}
\vspace{-0.1cm}
\end{table}

\section{CONCLUSIONS}
\label{sec:conclusions}
In this work, we leverage the SimulMT task to improve the performance of SimulST system. Various techniques from the offline ST domain, such as online KD and CAR are found to be beneficial for the SimulST task. To improve the performance further, we also propose \textbf{Decision Attentive Regularization}. It improves the \textit{read/write} decision policy for SimulST by using the monotonic attention energies of the SimulMT model. This work improves the performance of MMA-based SimulST by $35\%$ or 4.5 BLEU points across different latency regimes. 

% \clearpage
\bibliographystyle{IEEEtran}

\bibliography{mybib}

% \begin{thebibliography}{9}
% \bibitem[1]{Davis80-COP}
%   S.\ B.\ Davis and P.\ Mermelstein,
%   ``Comparison of parametric representation for monosyllabic word recognition in continuously spoken sentences,''
%   \textit{IEEE Transactions on Acoustics, Speech and Signal Processing}, vol.~28, no.~4, pp.~357--366, 1980.
% \bibitem[2]{Rabiner89-ATO}
%   L.\ R.\ Rabiner,
%   ``A tutorial on hidden Markov models and selected applications in speech recognition,''
%   \textit{Proceedings of the IEEE}, vol.~77, no.~2, pp.~257-286, 1989.
% \bibitem[3]{Hastie09-TEO}
%   T.\ Hastie, R.\ Tibshirani, and J.\ Friedman,
%   \textit{The Elements of Statistical Learning -- Data Mining, Inference, and Prediction}.
%   New York: Springer, 2009.
% \bibitem[4]{YourName17-XXX}
%   F.\ Lastname1, F.\ Lastname2, and F.\ Lastname3,
%   ``Title of your INTERSPEECH 2022 publication,''
%   in \textit{Interspeech 2022 -- 23\textsuperscript{rd} Annual Conference of the International Speech Communication Association, September 18-22, Incheon, Korea, Proceedings, Proceedings}, 2022, pp.~100--104.
% \end{thebibliography}

\end{document}